\documentclass[preprint,12pt,authoryear]{elsarticle}

\usepackage{graphicx} 
\usepackage{amssymb}
\usepackage{amsmath}
\usepackage{natbib}
\usepackage{booktabs} 
\usepackage{multirow}
\usepackage{url}
\usepackage{color}

\usepackage{float}
\newtheorem{definition}{Definition}
\newtheorem{thm}{Theorem}
\newtheorem{lem}{Lemma}
\newdefinition{rmk}{Remark}
\newproof{pf1}{Proof of $\mathbf{P1}$}
\newproof{pf2}{Proof of $\mathbf{P2}$}
\newproof{pf3}{Proof of $\mathbf{P3}$}
\newproof{pf4}{Proof of $\mathbf{P4}$}
\newproof{pot}{Proof of Theorem \ref{thm1}}
\newproof{pol1}{Proof of Lemma \ref{lem1}}
\newproof{pol2}{Proof of Lemma \ref{lem2}}

\title{Product Depth for Temporal Point Processes Observed Only Up to the First $k$ Events}
\author[1]{Chifeng Shen}
\author[1]{Yuejiao Fu\corref{cor2}}
\author[2]{Xiaoping Shi}
\author[1]{Michael Chen}
\ead{sen86@yorku.ca}
\cortext[cor2]{Correspondence: yuejiao@yorku.ca}

\affiliation[1]{organization={Department of Mathematics and Statistics},
addressline={York University},
city={Toronto},
postcode={ON M3J 1P3},
country={Canada}}

\affiliation[2]{organization={Department of Computer Science, Mathematics, Physics and Statistics},
addressline={University of British Columbia},
city={Kelowna},
postcode={BC V1V 1V7},
country={Canada}}

\begin{document}

\begin{abstract}

Temporal point processes (TPPs) model the timing of discrete events along a timeline and are widely used in fields such as neuroscience and finance. Statistical depth functions are powerful tools for analyzing centrality and ranking in multivariate and functional data, yet existing depth notions for TPPs remain limited. In this paper, we propose a novel product depth specifically designed for TPPs observed only up to the first $k$ events. Our depth function comprises two key components: a normalized marginal depth, which captures the temporal distribution of the final event, and a conditional depth, which characterizes the joint distribution of the preceding events. We establish its key theoretical properties and demonstrate its practical utility through simulation studies and real data applications.
\end{abstract}


\maketitle

\section{Introduction}
The temporal point process (TPP) is a random process whose realizations consist of the times of discrete events localized along a real timeline \citep{introductionTP}. 
The study of TPPs dates back to the early twentieth century, with applications in modeling insurance claims and telephone traffic \citep{brockmeyer1948life,Cramr1969HistoricalRO}. 
Today, temporal point pattern data are widely observed across diverse fields such as neuroscience \citep{brown1998statistical,kass2001spike} and finance \citep{hawkes2018hawkes}. 

In this work, we aim to investigate the distributional characteristics of TPPs---such as centrality and ranking---which present unique challenges due to their temporally ordered nature. Statistical depth, which quantifies the center-outward rank of a sample relative to its underlying distribution, has emerged as a powerful tool in nonparametric inference. It extends classical univariate concepts such as the median, quantiles, ranks, and outlyingness to multivariate settings. Various notions of depth have been proposed in the literature. \citet{tukey1975mathematics} was the first to systematically introduce the notion of statistical depth for multivariate data. Subsequent developments include the simplicial depth by \citet{liu1990notion}, Mahalanobis Depth by \citet{liu1993quality}, and likelihood depth by \citet{fraiman1999multivariate}. \citet{zuo2000general} provided a unified framework for multivariate depth functions by identifying four desirable properties: affine invariance, maximality at the center, monotonicity relative to the center, and vanishing at infinity. 

This work develops a depth-based statistical model for temporal point processes observed only through their first $k$ events. The observation window is unbounded, in the sense that the time of the $k$th event is unrestricted, and the methodology is designed to accommodate this open-ended temporal domain. A motivating example is the cell-division data obtained from an automatic embryo-monitoring system \citep{shixiaopin}. This particular dataset consists of 100 embryo trajectories, each recording the time of three division events, denoted by $(S_1, S_2, S_3)$, where the third event time $S_3$ is unrestricted. Another example arises from high-performance sports science: the 40m sprint-test data provided by \citet{sport}. Focusing on athletes' performance, the recorded times at which each athlete reaches 10m, 20m, 30m, and 40m are denoted by $S_1, S_2, S_3,$ and $S_4$, respectively, with the final time $S_4$ similarly unrestricted. Both datasets exemplify the type of partially observed but temporally open-ended point process trajectories that motivate our proposed framework.

In the existing literature, statistical depths specifically designed for temporal point processes are rather limited. One of the earliest related approaches is the generalized Mahalanobis depth proposed by \citet{liu2017generalized}. For TPP data with a variable number of events within a time domain, the event distribution can be treated within a multivariate framework. Building on this idea, \citet{liu2017generalized} proposed a generalized Mahalanobis depth consisting of two components: 
(1) a weighted depth for event counts, and (2) a conditional depth for the timing of events, given a fixed number of events within the time domain. However, this approach fails to capture the intrinsic temporal ordering of events, which is fundamental to TPPs. More recent developments, such as the Dirichlet depth \citep{qi} and the ILR depth \citep{ilr}, provide novel frameworks for ranking TPPs defined on a fixed time domain. Neither method can be directly applied to typical data sets—such as the cell-division data introduced earlier—because those data are observed on an unbounded time domain.

Motivated by the frameworks of \citet{liu2017generalized} and \citet{qi}, we propose a novel product depth specifically designed for temporal point processes observed only up to the first $k$ events. The proposed depth integrates marginal and conditional event-time distributions, allowing for a refined quantification of centrality and the detection of anomalies within the process. 
We establish the key theoretical properties of the proposed depth and demonstrate its effectiveness through simulation studies and real-data applications.

The remainder of the paper is organized as follows. 
Section~2 introduces the proposed depth function and discusses its theoretical properties. 
Section~3 presents simulation studies to assess the ranking performance of the proposed method. 
Section~4 applies the method to real-world datasets. 
Finally, Section~5 concludes with a summary and discussion.

\section{Methods}
When a temporal point process is observed only up to its first
$k$ events, assessing centrality requires a tailored notion of depth. To this end, we propose Product Depth, together with its theoretical properties, detailed in the sections that follow.

\subsection{Depth function}

Let $T_0 \geq 0$ be the starting time of the TPPs.  For any $ k \in \mathbb{Z}^{+}$ , let $\mathbb{S}_k=\{(s_1,s_2,\cdots,s_k)\in \mathbb{R}^k|T_0 \leq s_1 \leq s_2 \leq \cdots \leq s_k\}$ denote the set of all point processes with cardinality $k$. The boundary set is defined as $\mathbb{B}_k=\{(s_1,s_2,\cdots,s_k)\in \mathbb{S}_k|\textit{at least one equality holds in } T_0 \leq s_1 \leq s_2 \leq \cdots \leq s_k \}.$ A depth function is a map $D: \mathbb{S}_k \to \mathbb{R}^{+}.$ 

A point process observed only up to its first $k$ events provides two pieces of information: the marginal distribution of the 
$k$th event time and the conditional distribution of the preceding event times. To utilize the information, the depth is defined as a product of the following two terms: (1) a normalized one-dimensional depth of the $k$th event time, and (2) a multivariate depth of the preceding event times conditioned on the $k$th  event time. In light of  \cite{qi}, the formal definition of the depth function is given as follows.

\begin{definition}
Given a random point process $\boldsymbol{S}_k\in \mathbb{S}_k$ w.r.t. a probability measure $P_{\mathbf{S}_k}$, and a start point $s_0=T_0$. Denote $P_{S_k}$ as a probability measure on $S_k$, and $P_{\boldsymbol{S}_k|S_k}$ as the probability measure on the $\boldsymbol{S}_k$, conditioned on $S_k$.
For a realization $\boldsymbol{s}_k \in \mathbf{S}_k$, the depth $D(\boldsymbol{s}_k;P_k)$ is defined as:
\begin{equation}\label{eq1}
D(\boldsymbol{s}_k;P_{\mathbf{S}_k})=\omega (s_k;P_{S_k})^{\frac{|s_k-\eta|}{M-s_0}}D_c(\boldsymbol{s}_k;P_{\boldsymbol{S}_k|S_k}),
\end{equation}
where $\omega (s_k;P_{S_k})=\frac{D_1(s_k;P_{S_k})}{\underset{x}{\max}D_1(x;P_{S_k})}$ is the normalized one dimensional depth, $\eta=\underset{t}{\arg\max}\ \big(\omega(t;P_{S_k})\big)$, $M$ is a non-negative hyperparameter, and $D_c(\boldsymbol{s}_k;P_{\boldsymbol{S}_k|S_k})$ is the depth of $\boldsymbol{s}_k$ conditioned on $S_k$.
\end{definition}

\begin{rmk}
The formulation in Definition \ref{eq1} gives the general structure of the depth 
based on a marginal depth $\omega (s_k;P_{S_k})$ and a conditional depth 
$D_c(\boldsymbol{s}_k;P_{\boldsymbol{S}_k|S_k})$. 
In this paper, the definitions of $\omega (s_k;P_{S_k})$ and $D_c(\boldsymbol{s}_k;P_{\boldsymbol{S}_k|S_k})$, as well as the hyperparameter $M$ used in this paper, are discussed below. 
\end{rmk}

In Equation (\ref{eq1}), the marginal depth $\omega (s_k;P_{S_k})$ utilizes the information on the last event time $S_k$. There are various methods that can be used to define $\omega (s_k;P_{S_k})$, and a one-dimensional Mahalanobis Depth $D_1(\cdot)$ is adopted, and $M = \mathbb{E}(S_k)$ here. 
\begin{equation*}
    D_1(s_k;P_{S_k})=[1+\frac{(s_k-s_0-\mu_{S_k-S_0})^2}{\sigma^2_{S_k-S_0}}]^{-1},
\end{equation*}
Where $\mu_{S_k-S_0}$ and $\sigma^2_{S_k-S_0}$ are the mean and variance of $S_k-S_0$ respectively. For the conditional depth $D_c(\boldsymbol{s}_k;P_{\boldsymbol{S}_k|S_k})$, the following depth is adopted. 
\begin{eqnarray*} 
    D_c(\boldsymbol{s}_k;P_{\boldsymbol{S}_k|S_k}) 
    &=&  \prod_{i=1}^{k}\big[\frac{s_i-s_{i-1}}{(s_k-s_0)\bar{\mu}_i}\big]^{\bar{\mu}_i},
\end{eqnarray*}
where $\bar{\mu}_i=E(\frac{S_i-S_{i-1}}{S_k-S_0})$, and $S_0=T_0$. 

For samples drawn from a  point process observed only up to the first $k$ events, the empirical method can be used to estimate $\bar{\mathbb{\mu}}_k=(\bar{\mu}_1,\bar{\mu}_2,\cdots,\bar{\mu}_k)$, and $D_1(s_k;P_{S_k})$. Accordingly, the sample version of $D_1(s_k;P_{S_k})$ and $D_c(\mathbf{s}_k;P_{\mathbf{S}_k|S_k})$ are defined as follows.

\begin{definition}\label{def2}
Given a sequence of point process realizations observed only up to the first $k$ events, $\boldsymbol{s}_k=(s_1,s_2,\cdots,s_k) \in \mathbb{S}_k$, and a start point $s_0=T_0$. Assume $\bar{\boldsymbol{u}}_k=(\bar{u}_1,\bar{u}_1,,\cdots,\bar{u}_k)$ to be the estimated sample means of unit inter-event times $\frac{\boldsymbol{U}_k}{S_K-S_0}$, the sample version of $D_1$ and $D_c$ in Equation (\ref{eq1}) can be defined as
\begin{equation}\label{eq2}   
D_1(s_k;P_{S_k}^{(n)})=[1+\frac{(s_k-s_0-\bar{S}_k)^2}{S^2}]^{-1},
\end{equation}
\begin{equation}\label{eq3}    D_c(\boldsymbol{s}_k;P_{\boldsymbol{S}_k|S_k}^{(n)})=\prod_{i=1}^{k}\Big[\frac{s_i-s_{i-1}}{(s_k-s_0)\bar{u}_i}\Big]^{\bar{u}_i},
\end{equation}
where $n$ is the sample size, $\bar{S}_k$ and $S^2$ are the sample mean and variance of $S_k-S_0$, and $P_{S_k}^{(n)}$ and $P_{\boldsymbol{S}_k|S_k}^{(n)}$ are the empirical distributions.
\end{definition}

\begin{rmk}
For the Homogeneous Poisson Process (HPP)  observed only up to its first $k$ events, $\boldsymbol{S}_k=(S_1,S_2,\cdots,S_k) \in \mathbb{S}_k$ with constant $\lambda$, the last event time $S_k$ follows an Erlang distribution with parameters $k$ and $\lambda$. Denote $\boldsymbol{U}_k=(U_1,U_2,\cdots,U_k)$, where $U_i=S_i-S_{i-1}$. Since $\frac{S_i-S_{i-1}}{S_K-S_0}$ and $S_k$ are independent $\forall i=1,2,\cdots,k$, and $\frac{\boldsymbol{U}_k}{S_K-S_0}|S_k$ follows the Dirichlet distribution with parameters $(\alpha_i=1,i=1,2,\cdots,k)$,  $\boldsymbol{\bar{\mu}}_k=E(\frac{\boldsymbol{U}_k}{S_K-S_0})=E(\frac{\boldsymbol{U}_k}{S_K-S_0}|S_k)$ is $(\frac{1}{k},\frac{1}{k},\cdots,\frac{1}{k})$. The conditional depth can be modified as follows.    
\end{rmk}

\begin{definition}
For an observed HPP observed only up to its first $k$ events, $\boldsymbol{s}_k=(s_1,s_2,\cdots,s_k)$, with start time $s_0=T_0$, the conditional depth is defined as
\begin{eqnarray}\label{eq4}
D_c(\boldsymbol{s}_k;P_{\boldsymbol{S}_k|S_k})&=&  k\prod_{i=1}^{k}\Big(\frac{s_i-s_{i-1}}{s_k-s_0}\Big)^{\frac{1}{k}}.
\end{eqnarray}
\end{definition}

\subsection{Properties}
The important properties for depth function on multivariate data are summarized by \cite{zuo2000general}. \cite{qi} further discussed desirable properties for a conditional depth function for a point process. Motivated by these works, we outline and discuss four desirable properties below.
\begin{itemize}
    \item[$\mathbf{P1}$] \textit{Continuity and vanishing at infinity and boundary}: Since the event time is continuous, the continuity of the depth is necessary. In addition, the depth of a point $x$ should approach zero as $||\mathbf{s}_k|| \to \infty$, and it is zero when $\mathbf{s}_k \in \mathbb{B}_k, \textit{the boundary set}.$ Here and throughout, $\|\cdot\|$ denotes the $L_2$ (Euclidean) norm.
    \item[$\mathbf{P2}$] \textit{Maximality at the center}: It is natural, since the center should have the maximum depth value. 
    \item[$\mathbf{P3}$] \textit{Scale and shift invariance}: It is a special case of the affine invariance property. Shift and scaling the underlying measurements should not affect the depth value of $\mathbf{s}_k$  
    \item[$\mathbf{P4}$] \textit{Monotonicity relative to the deepest point}: the depth value of $\mathbf{s}_k$ decreases monotonically as it moves away from the center along any fixed ray through the center.     
\end{itemize}

As noted by \cite{qi}, traditional multivariate notions of the center cannot be directly applied to temporal point processes because of their inherent temporal ordering and scale structure. For point processes observed only up to their first $k$ events with start time $S_0=T_0$, we define the central event-time vector as $\Theta_k= (\theta_1,\theta_2,\cdots,\theta_k)$, where $\theta_i=S_0+\mathbb{E}(S_k-S_0)\mathbb{E}\left(\sum^{i}_{j=1}\frac{S_j-S_{j-1}}{S_k-S_0}\right)$. This definition combines the average duration $\mathbb{E}(S_k-S_0)$ with the expected relative timing pattern $\mathbb{E}\left(\frac{S_i-S_{i-1}}{S_k-S_0}\right)$, thus capturing both the scale and shape characteristics of the process.

In the special case where $\mathbf{S}_k/S_k$ and $S_k$ are independent, 
such as for homogeneous Poisson processes (HPP), the center reduces to $\mathbb{E}(\mathbf{S}_k)$. 
The main theoretical results are summarized in Theorem~\ref{thm1}.

\begin{thm}\label{thm1}
Let $\mathbb{P}$ be the class of distributions on $\mathbb{S}_k$ and $P_{\mathbf{s}_k}$ be the distribution of a given random point process $\mathbf{S}_k \in \mathbb{S}_k$.
Assume the mapping $D(\cdot;\cdot): \mathbb{S}_k \times \mathbb{P} \to \mathbb{R}^{+}$ be bounded, and $\Theta_k$ is adopted as a measurement of centrality, where $\Theta_k= (\theta_1,\theta_2,\cdots,\theta_k)$, and $\theta_i=S_0+\mathbb{E}(S_k-S_0)\mathbb{E}\left(\sum^{i}_{j=1}\frac{S_j-S_{j-1}}{S_k-S_0}\right)$, $\forall i=1,2,\cdots,k$. The Product Depth satisfies the following 4 properties:
\begin{itemize}
\item[(i)] $D(\mathbf{s}_k;P_{\mathbf{S}_k})$ is a continuous map from $\mathbb{S}_k$ to $\mathbb{R}^{+}$, and $D(\mathbf{s}_k;P_{\mathbf{S}_k})\to 0$ as $||\mathbf{s}_k|| \to \infty$, and  $D(\mathbf{s}_k;P_{\mathbf{S}_k})=0, \forall \ \mathbf{s}_k \in \mathbb{B}_k$. $\|\cdot\|$ denotes the $L_2$ (Euclidean) norm;
\item[(ii)]    $D(\Theta_k;P_{\mathbf{S}_k})=\underset{\mathbf{s}_k \in \mathbb{S}_k}{\sup}D(\mathbf{s}_k;P_{\mathbf{S}_k})$ holds $\forall \ P_{\mathbf{S}_k} \in \mathbb{P}$ having central point $\Theta_k$;
\item[(iii)]
$\forall \ a \in \mathbb{R}^{+}$ and $\forall \  b \in \mathbb{R}$, $D(\mathbf{s}_k;P_{\mathbf{S}_k})=D(a\mathbf{s}_k+b;P_{a\mathbf{S}_k+b});$ and
\item[(iv)]
$\forall \ P_{\mathbf{S}_k} \in \mathbb{P}$ having central point $\Theta_k,$ $D(\mathbf{s}_k;P_{\mathbf{S}_k}) \leq D\big(\Theta_k+\alpha(\mathbf{s}_k-\Theta_k);P_{\mathbf{S}_k}\big), \forall \mathbf{s}_k \in \mathbb{S}_k, \alpha \in [0,1].$   
\end{itemize} 
\end{thm}
The details of the proof are given in the appendix.
\begin{rmk}
    Compared with the properties proposed by \cite{zuo2000general} and \cite{qi}, the main distinction lies in Property P-1. The property proposed by \cite{zuo2000general} does not consider vanishing at the boundary, which is important for capturing the natural ordering inherent in temporal point processes. In contrast, vanishing at infinity is not considered a property by \cite{qi}, since the time domain is assumed to be fixed in their framework.  In our setting - temporal point processes with a fixed number of events and an unbounded time domain — both vanishing at the boundary and at infinity are essential. This motivates a refinement of Property P-1 to incorporate both conditions. 
\end{rmk}

\section{Simulation Study}
In this section, we evaluate the ranking performance of the Product Depth on both HPP data and state-dependent point process data, and compare it with that of the general Mahalanobis Depth.

\begin{figure}[H]
    \centering
    \includegraphics[width=1\linewidth]{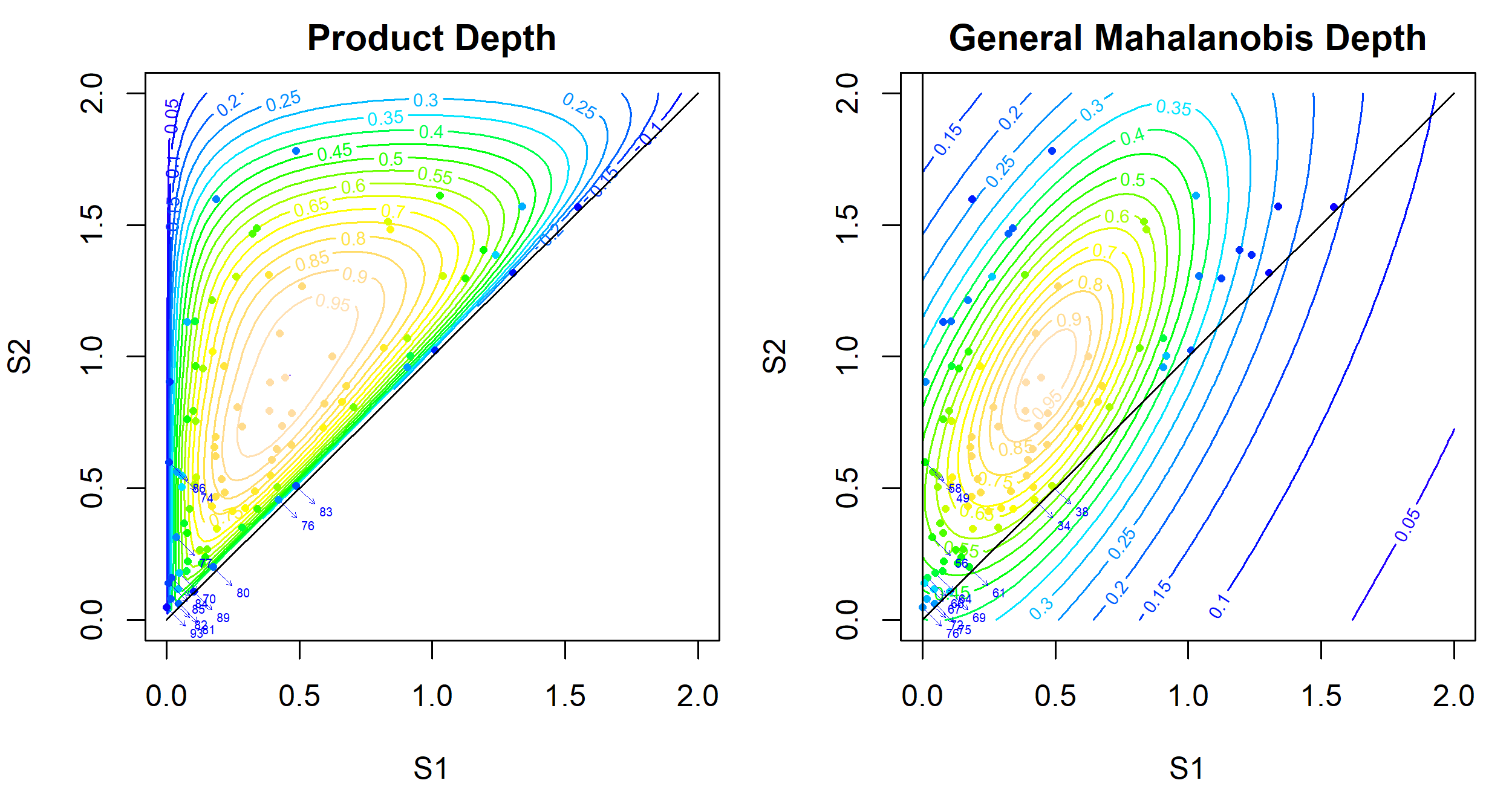}
    \caption{The contour curves of the Product Depth and the general Mahalanobis Depth in HPP, where $\lambda=2$ and $k=2$; The blue number indicates the rank of the point based on the corresponding depth measure; The ranks of the points are 76, 83, 74, 77, 86, 80, 70, 84, 85, 89, 82, 81, 93 (left plot), and 34, 38, 49, 56, 58, 61, 64, 65, 67, 69, 72, 75, 76 (right plot).}
    \label{fig:1}
\end{figure}

We begin with Case I, which consists of 100 realizations from a homogeneous Poisson process (HPP) observed only up to the first $k$ events with $\lambda=2$. The contour plots for the general Mahalanobis Depth and the Product Depth (Equation (\ref{eq1})) are illustrated in Figure \ref{fig:1}. For the Product Depth, we apply the marginal depth from Equation (\ref{eq2}) and the conditional depth from Equation (\ref{eq4}). Compared to the general Mahalanobis Depth, the Product Depth ranks the realizations differently, particularly for points near the boundary. As we mentioned in the introduction, the temporal point process data are ordered, meaning that in this case, $T_0 < S_1 <S_2$. The points are confined to a region shaped as an upper triangle. Compared to the elliptical contour of the general Mahalanobis Depth, the triangle-like contour of the Product Depth is more appropriate within the restricted region. In Figure \ref{fig:1}, we highlight the ranks of several points near the boundary using both depth measures. The ranks of the points are 34, 38, 49, 56, 58, 61, 64, 65, 67, 69, 72, 75, 76 in the general Mahalanobis Depth contour, and 76, 83, 74, 77, 86, 80,70, 84, 85, 89, 82, 81, 93 in the Product Depth contour. Compared with the general Mahalanobis Depth, Product Depth assigns a relatively smaller depth value to the points near the boundary and zero to those on the boundary, resulting in higher ranks.   

\begin{figure}[H]
    \centering
    \includegraphics[width=1\linewidth]{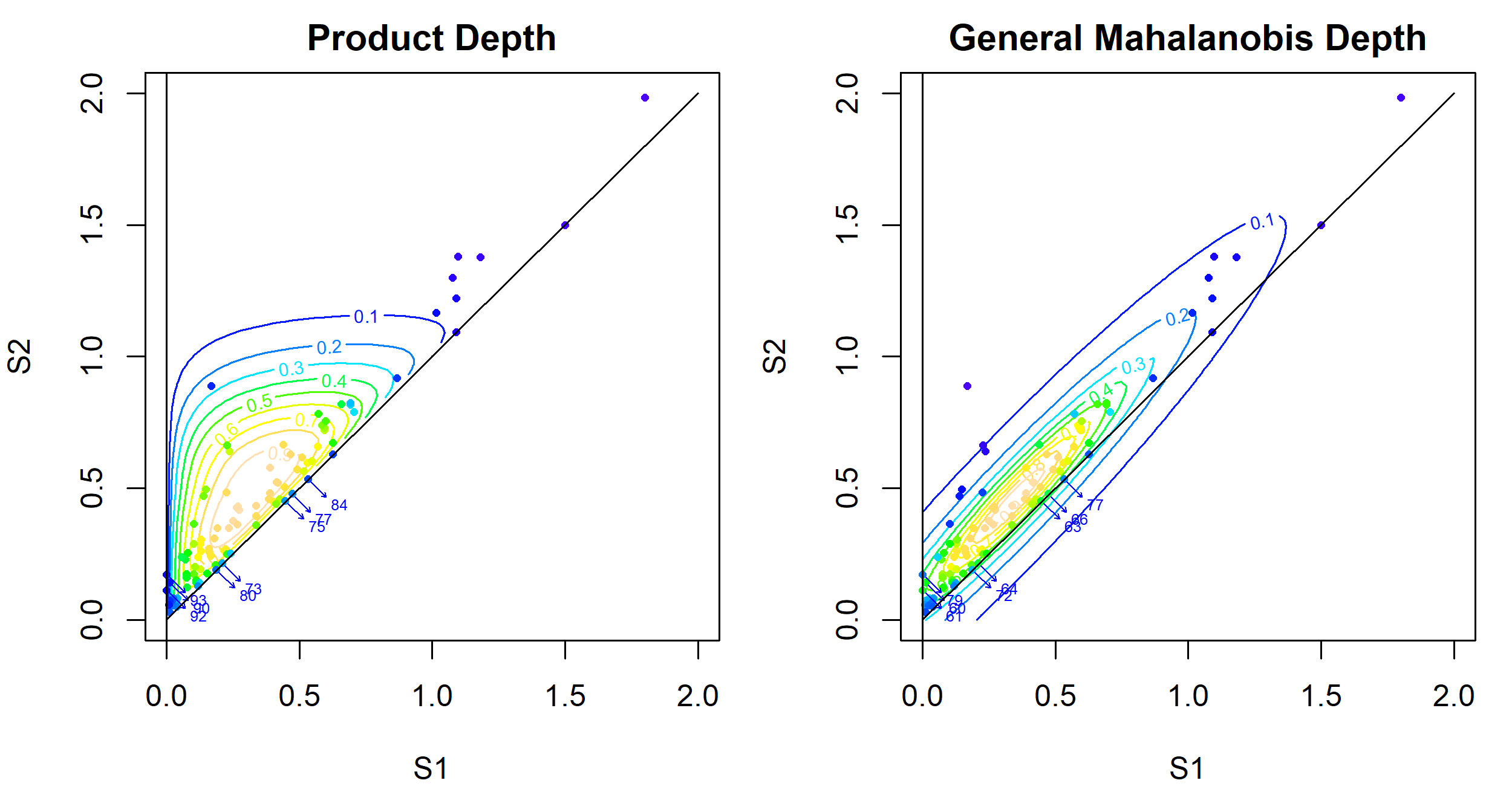}
    \caption{The contour curves of the Product Depth and the general Mahalanobis Depth, where $(\lambda_1,\lambda_2) =(2.5,10)$ and $k=2$; The blue number indicates the rank of the point based on the corresponding depth measure; The ranks of the points are 60, 61, 63, 64, 66, 72, 77, 79 (left plot), and 90, 92, 75, 73, 77, 80, 84, 93 (right plot).}
    \label{fig:2}
\end{figure}

Next, we consider a Poisson point process observed only up to their first $k$ events, with a state-dependent (piecewise-constant) conditional intensity. In this setting, the instantaneous rate of event occurrence depends on the number of events that have already occurred by time~$t$, that is,
$$
\lambda(t \mid \mathcal{H}_t)=
\begin{cases}
\lambda_1, & \text{if } N(t)=0,\\[4pt]
\lambda_2, & \text{if } N(t)=1.
\end{cases}
$$
Here, $\mathcal{H}_t$ denotes the history of the process up to time $t$. Under this specification, the inter-event durations are independent exponentials:
$$
U_1 = S_1 - S_0 \sim \mathrm{Exp}(\lambda_1), \qquad
U_2 = S_2 - S_1 \sim \mathrm{Exp}(\lambda_2), \qquad
U_1 \perp U_2.
$$
For this case, we set $(\lambda_1,\lambda_2)=(2.5,10)$ and generate 100 realizations.
This configuration reflects a process whose first event occurs under a slower rate, 
followed by a substantially faster second event, resulting in a nonstationary temporal structure 
that is not determined by a fixed observation window. Figure \ref{fig:2} presents the contour curves based on general Mahalanobis Depth and the Product Depth (Equation (\ref{eq1})). In this case, marginal depth (Equation (\ref{eq2})) and condition depth (Equation (\ref{eq3})) are applied in the computation of the Product Depth. Similar to the previous cases, the ranks of several points near the boundary are highlighted. The ranks of the selected points
are 60, 61, 63, 64, 66, 72, 77, and 79 in the general Mahalanobis Depth contour, and 90, 92, 75, 73, 77, 80, 84, and 93 in the Product Depth contour. The triangle-shaped contour curve is more compatible with the restricted triangular region, leading to a more appropriate ranking.

\begin{figure}[H]
    \centering    \includegraphics[width=1\linewidth]{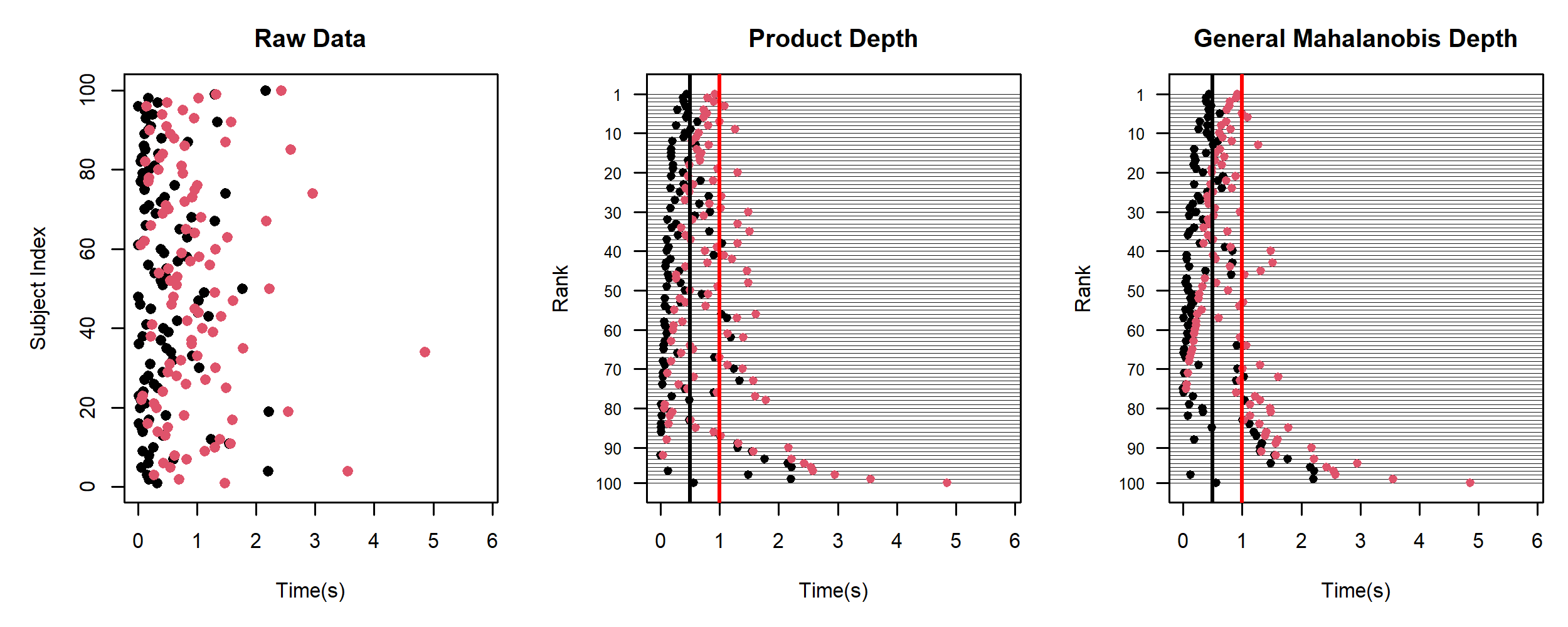}
    \caption{Left: Scatter plot of the HPP, where $\lambda=2$ and $k=2$; Middle: Scatter plot after ranking based on Product Depth; Scatter plot after ranking based on general Mahalanobis Depth. In all plots, black points correspond to the first event time $S_1$, and red points correspond to the second event time $S_2$. The black and red vertical lines represent the means of $S_1$ and $S_2$, respectively.}
    \label{fig:3}
\end{figure}

\begin{figure}[H]
    \centering    \includegraphics[width=1\linewidth]{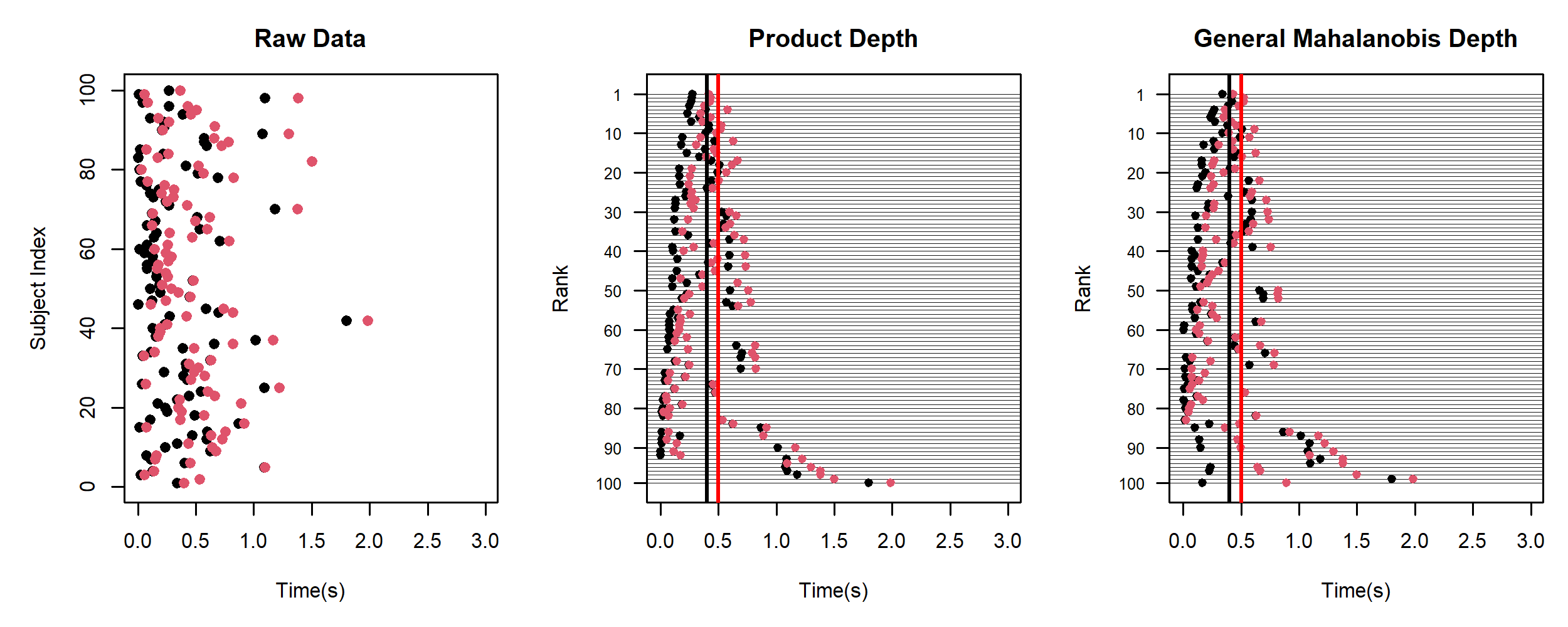}
    \caption{Left: Scatter plot of the data, where $(\lambda_1,\lambda_2) =(2.5,10)$ and $k=2$; Middle: Scatter plot after ranking based on Product Depth; Scatter plot after ranking based on general Mahalanobis Depth. In all plots, black points correspond to the first event time $S_1$, and red points correspond 
    to the second event time $S_2$. The black and red vertical lines represent the means of $S_1$ and $S_2$, respectively.}
    \label{fig:4}
\end{figure}
Figure \ref{fig:3} and Figure \ref{fig:4} present scatter plots of the ranked data in Case I and Case II, respectively. The black and red vertical lines represent the means of $S_1$ and $S_2$, respectively. Compared with the general Mahalanobis Depth, the Product Depth yields more symmetric rankings in both cases. This is primarily because points near the boundary $s_1=s_2$ tend to receive higher ranks under the general Mahalanobis Depth.

\section{Application}
\subsection{Cell Division}
\begin{figure}[H]
    \centering     \includegraphics[width=0.8\linewidth]{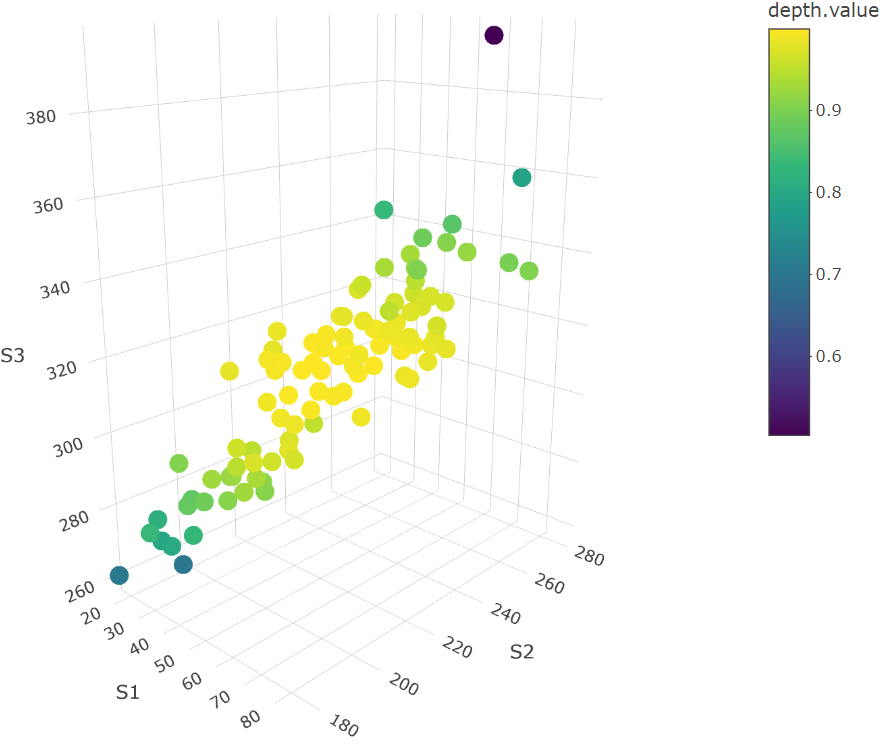}
      \caption{Cells with their depth value indicated by the different levels of color}
      \label{fig:7}
\end{figure}

In this section, we begin by applying the Product Depth to the Cell Division data. \cite{Cicconet2014} developed an automated embryo monitoring system using time-lapse imaging to record the timing of cell divisions. Figure \ref{fig:7} presents sample frames from a specific mouse embryo.
\cite{shixiaopin} formulated the detection of cell divisions as a multiple change-point problem, as the distribution of pixel values in image frames changes when a cell undergoes division. Their method developed a Bayesian-type statistic based on the shortest Hamiltonian path (SHP), combined with a ratio cut algorithm to accurately estimate change-point locations.

\begin{figure}[H]
    \centering    \includegraphics[width=1\linewidth]{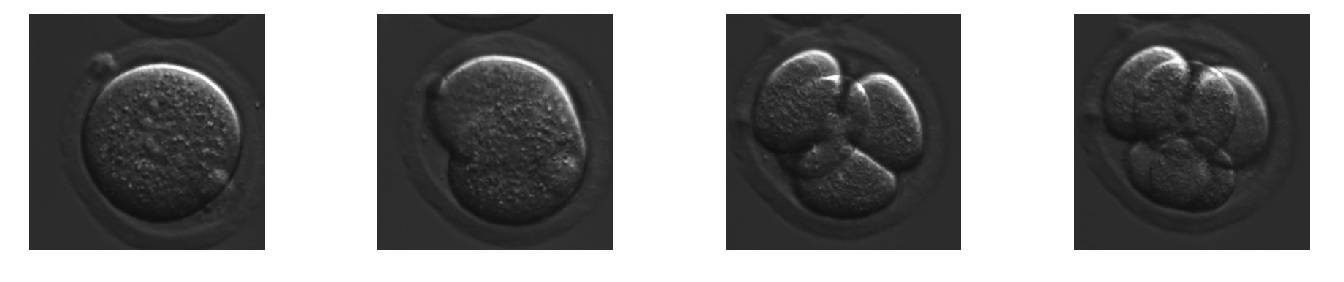}
    \caption{ The sample images, each of size $321 \times 321$, are taken from time points 1, 23, 195, and 259 in the folder E00 from celltracking.bio.nyu.edu. The first, second, and third cell division events occur at time points 22, 194, and 258, respectively.}
    \label{fig:5}
\end{figure}

\begin{figure}[H]
    \centering     \includegraphics[width=1\linewidth]{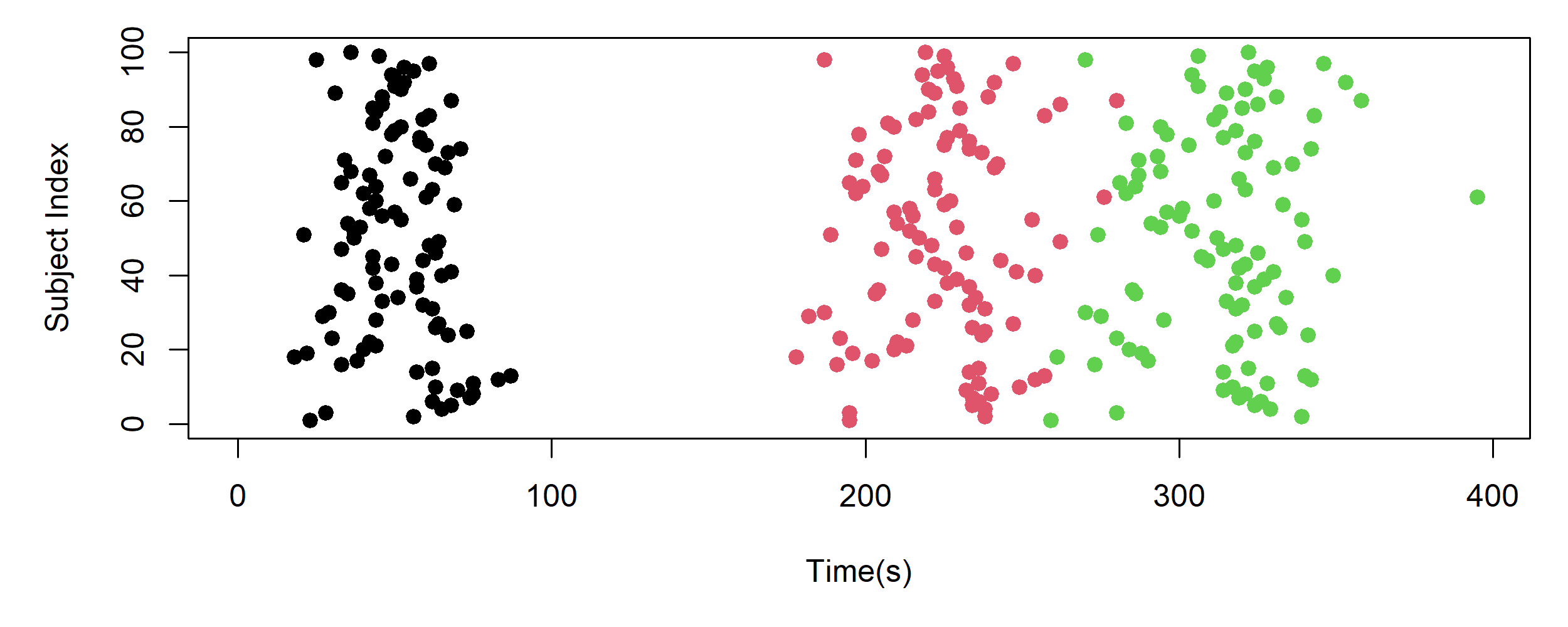}
      \caption{Cell Division Data; Black, red, and green points represent the first, second, and third division times ($S_1$, $S_2$, and $S_3$), respectively.}
      \label{fig:6}
\end{figure}

However, while \cite{shixiaopin} primarily focused on detecting unknown change-points, in many practical scenarios, the exact timing of events such as cell divisions is directly observed or precisely recorded. Motivated by this, our work shifts the analytical focus from event detection to characterizing the inherent structural features and centrality within temporal point processes (TPPs). To illustrate the utility of our proposed approach, we apply the depth function defined in Definition \ref{def2}. In the experiment, the cell division times are categorized as follows: $S_1$ denotes the first division time, $S_2$ corresponds to the second or third division time, and $S_3$ refers to the fourth or later division time. As shown in Figure \ref{fig:6}, the sample means of $S_1$, $S_2$, and $S_3$ are 50.64s, 223.84s, and 312.97s, respectively. These values suggest that the realizations are unlikely to be sampled from a homogeneous Poisson process (HPP).

In Figure \ref{fig:7}, the cells are visualized in a three-dimensional space, with their depth values estimated using the Product Depth. These depth values are represented using a 
topographic color scale, where lighter (yellow) colors indicate greater depth and darker (purple) colors indicate lower depth. Notably, the cells exhibit decreasing depth as they move outward from the central region, highlighting the ability of the Product Depth to capture centrality in a structured spatial pattern.

\begin{figure}[H]
    \centering     \includegraphics[width=1\linewidth]{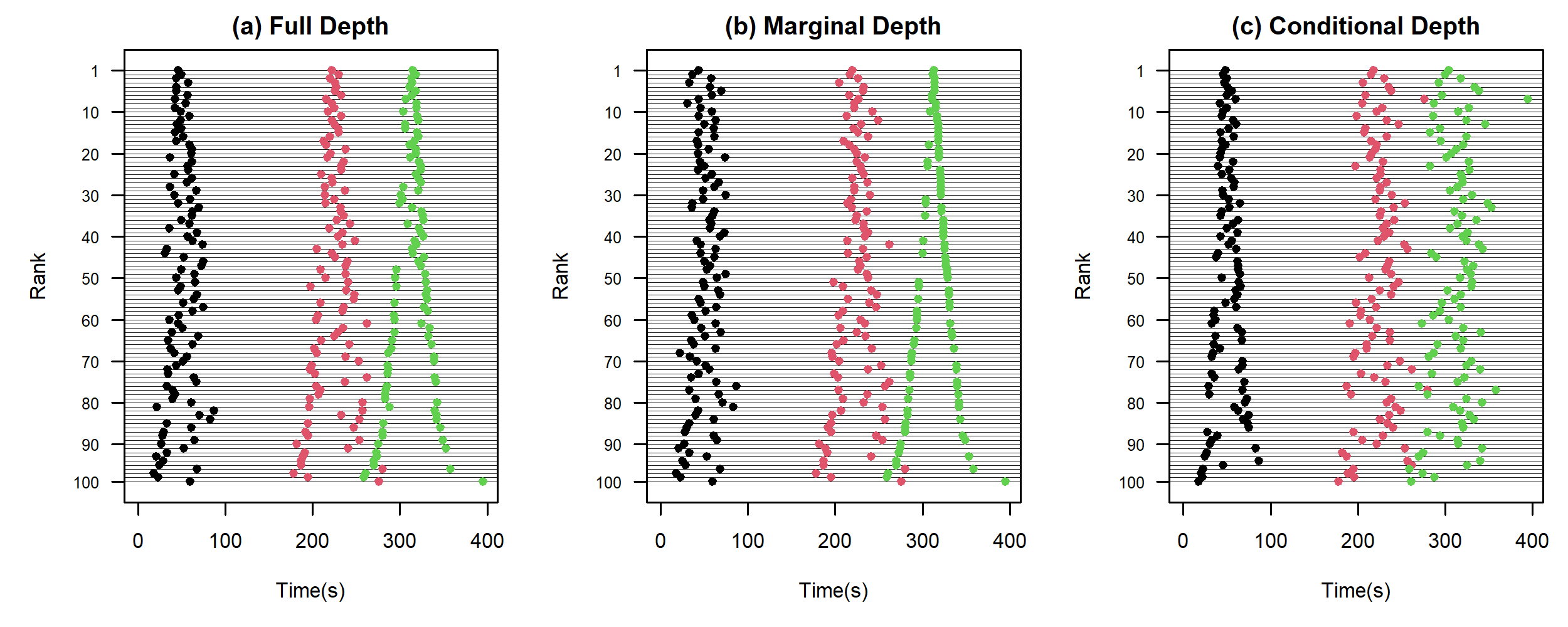}
      \caption{(a) Ranked cell based on whole Product Depth. (b) Ranked cell based on marginal depth. (c) Ranked cell based conditional depth. In all plots, black, red, and green points correspond to the first, second, and third event times ($S_1$, $S_2$, and $S_3$), respectively.}
      \label{fig:8}
\end{figure}

Based on the depth values, the cells can be ranked accordingly, as illustrated in Figure \ref{fig:8}(a). As expected, the marginal depth in Figure \ref{fig:8}(b) is driven by the final event time $S_3$, which reflects the overall temporal scale of the division process. It ranks cells according to how far their final division time deviates from the central timing. However, among cells with similar $S_3$, the marginal depth cannot distinguish differences in their intermediate event structure. This is particularly evident among the top-ranked cells, where the overall durations are similar but the relative spacing among $(S_1, S_2, S_3)$ varies. In contrast, the conditional depth in Figure \ref{fig:8}(c) focuses on ranking cells according to the relative spacing among $(S_1, S_2, S_3)$, regardless of the total duration. That is, the marginal depth orders cells from the center outward based on the final time (global scale), whereas the conditional depth orders them based on their internal timing pattern (relative structure). Figure \ref{fig:8}(a) combines these two components: it preserves the global temporal pattern associated with $S_3$, while also ranking cells according to the coherence of their internal event spacing. As a result, the full depth captures both the overall scale and the relative structure of the event timing, yielding a ranking that reflects the complete temporal pattern of the cell cycle.

\begin{figure}[H]
    \centering     \includegraphics[width=1\linewidth]{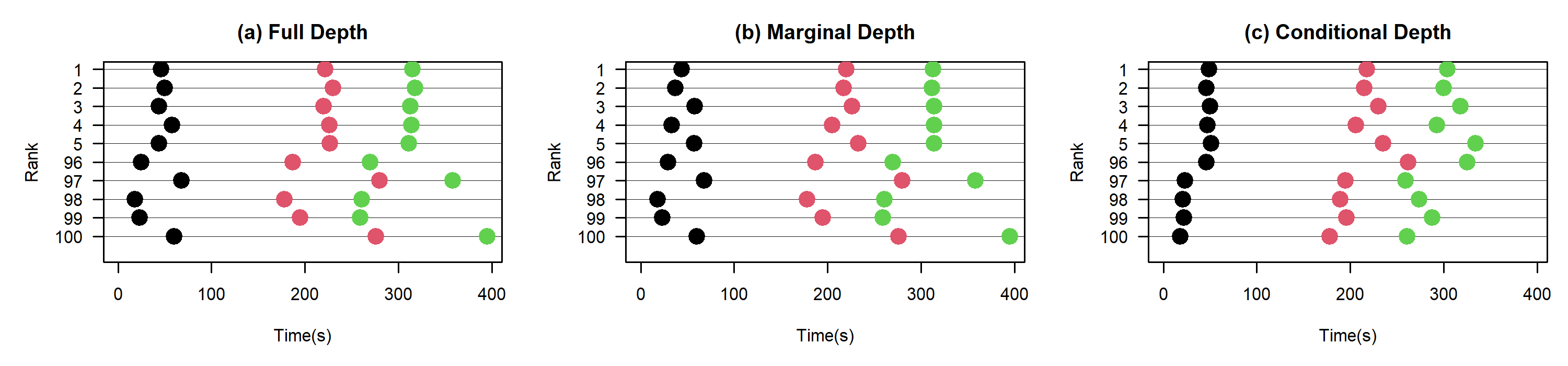}
      \caption{(a) The top 5 and bottom 6 ranked cell based on whole Product Depth. (b) The top 5 and bottom 6 ranked cell based on marginal depth. (c) The top 5 and bottom 6 ranked cell based conditional depth. In all plots, black, red, and green points correspond to the first, second, and third event times ($S_1$, $S_2$, and $S_3$), respectively.}
      \label{fig:9}
\end{figure}

This feature is more clearly illustrated in Figure \ref{fig:9}, which displays the top 5 and bottom 6 ranked cells. Compared to the conditional depth, the marginal depth more effectively identifies the abnormal cell whose final division time is close to 400 seconds. However, when using only the conditional depth, cells with similar depth ranks tend to exhibit more consistent patterns. By integrating both perspectives, the overall depth successfully captures the outlier while also maintaining pattern consistency among cells with comparable ranks.

\begin{figure}[H]
    \centering     \includegraphics[width=1\linewidth]{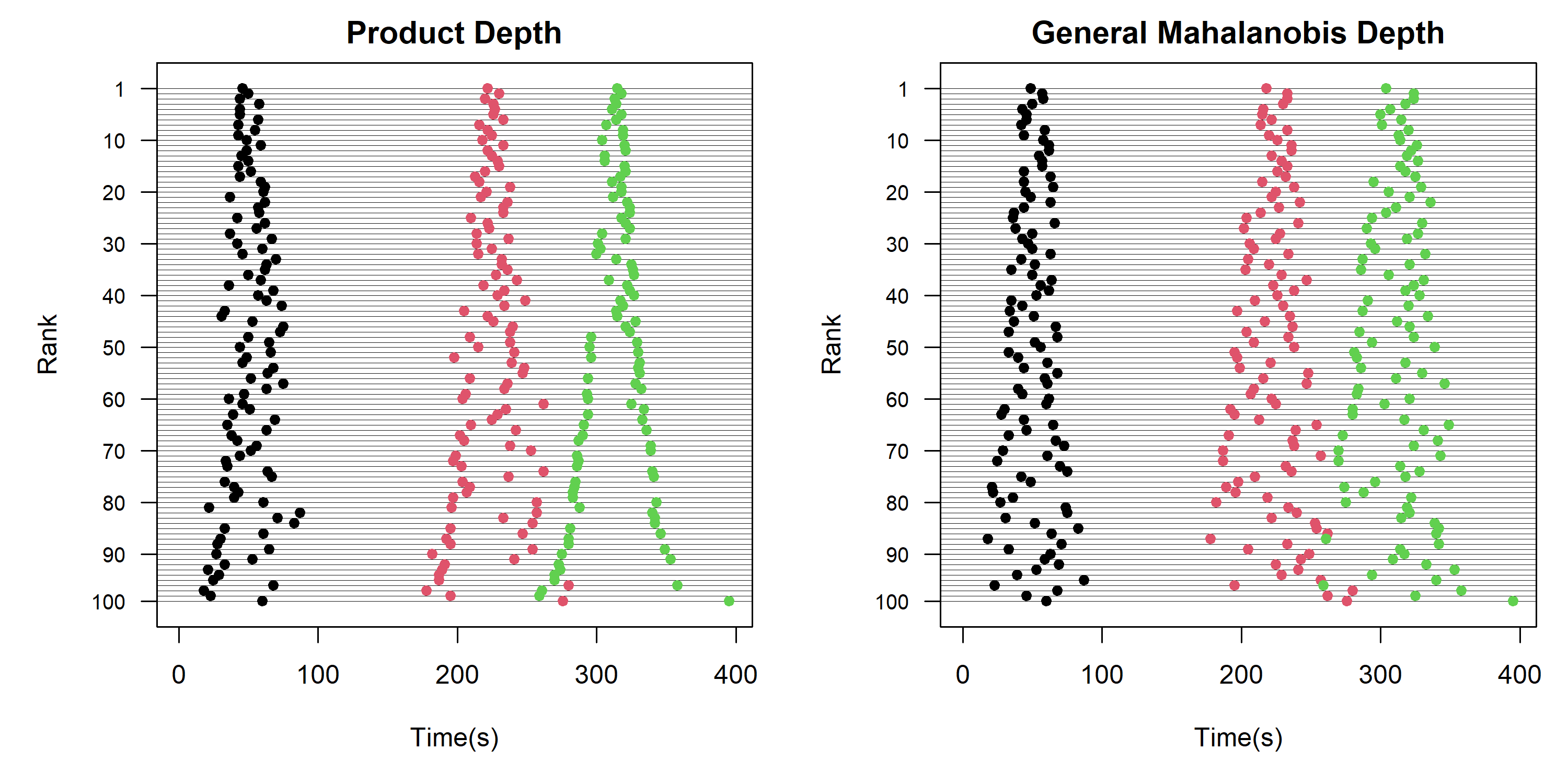}
      \caption{Left plot: Ranked cell based on Product Depth; Right plot: Ranked cell based on general Mahalanobis Depth. In all plots, black, red, and green points correspond to the first, second, and third event times ($S_1$, $S_2$, and $S_3$), respectively.} 
      \label{fig:10}
\end{figure}

Figure \ref{fig:10} presents the cell rankings based on Product Depth and general Mahalanobis Depth. Unlike the Mahalanobis Depth, the Product Depth yields a sharper and more well-defined peak in the rank plot, indicating a clearly identifiable center. This  structure suggests that the Product Depth is more sensitive to centrality in the data. Furthermore, the smooth and consistent ranking of ranks in the Product Depth plot reflects strong within-group cohesion, indicating a higher degree of within-group homogeneity compared to the more dispersed pattern observed under Mahalanobis Depth.

\subsection{40m Sprint tests}
\begin{figure}[H]
    \centering     \includegraphics[width=1\linewidth]{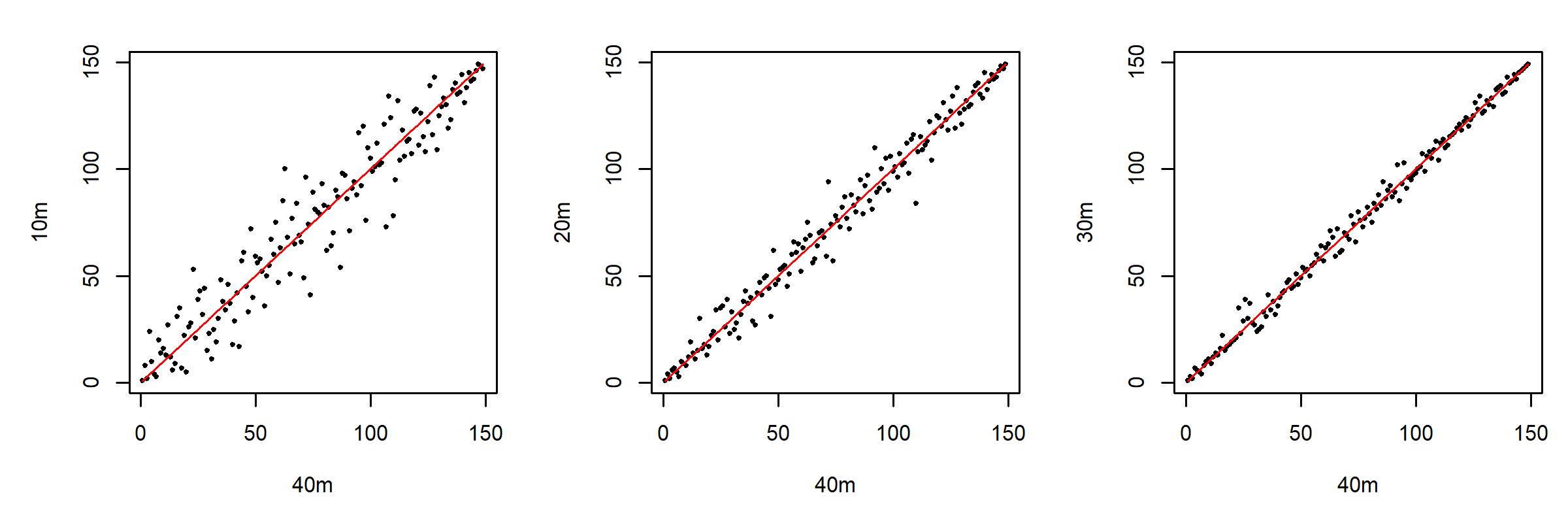}
      \caption{Left plot: athlete ranks based on recorded time of 10m versus 40m ; Middle plot: 20m versus 40m; Right Plot: 30m versus 40m. }
      \label{fig:11}
\end{figure}

Secondly, we compare the Product Depth and general Mahalanobis Depth using data from 40m sprint tests. The dataset, provided by \cite{sport}, includes athletes from various sports performing the 40m sprint under highly controlled conditions. For our analysis, we focus on soccer athletes, with the sample size of 150.  The recorded times at which each athlete reaches 10m, 20m, 30m, and 40m are denoted by $S_1, S_2, S_3,$ and $S_4$, respectively. Although the pairwise correlation coefficients among $S_1$, $S_2$, $S_3$, and $S_4$ are all close to 1, the athlete rankings based on the recorded times vary considerably. This is understandable, because the fastest athlete may not necessarily exhibit the greatest acceleration in the early stages. For each recorded split time, athletes can be ranked from fastest to slowest. Figure~\ref{fig:11} presents scatter plots of athlete rankings based on the times to reach 10m, 20m, 30m, and 40m, respectively. These plots reveal noticeable inconsistencies in rankings across different milestones, particularly between the initial 10m and the full 40m sprint. Each split captures a distinct phase of sprint performance: the 10m time reflects explosive acceleration and initial drive; the 10–20m segment measures continued acceleration; the 20–30m phase captures the transition toward maximal velocity; and the final 30–40m split assesses the athlete’s ability to reach or maintain top-end speed. Thus, to comprehensively evaluate an athlete’s multifaceted sprinting capability, it is essential to consider not only the final result but also the intermediate split times. The depth function offers a tool to distinguish between typical and atypical performance profiles from an overall perspective.

\begin{figure}[H]
    \centering     \includegraphics[width=1\linewidth]{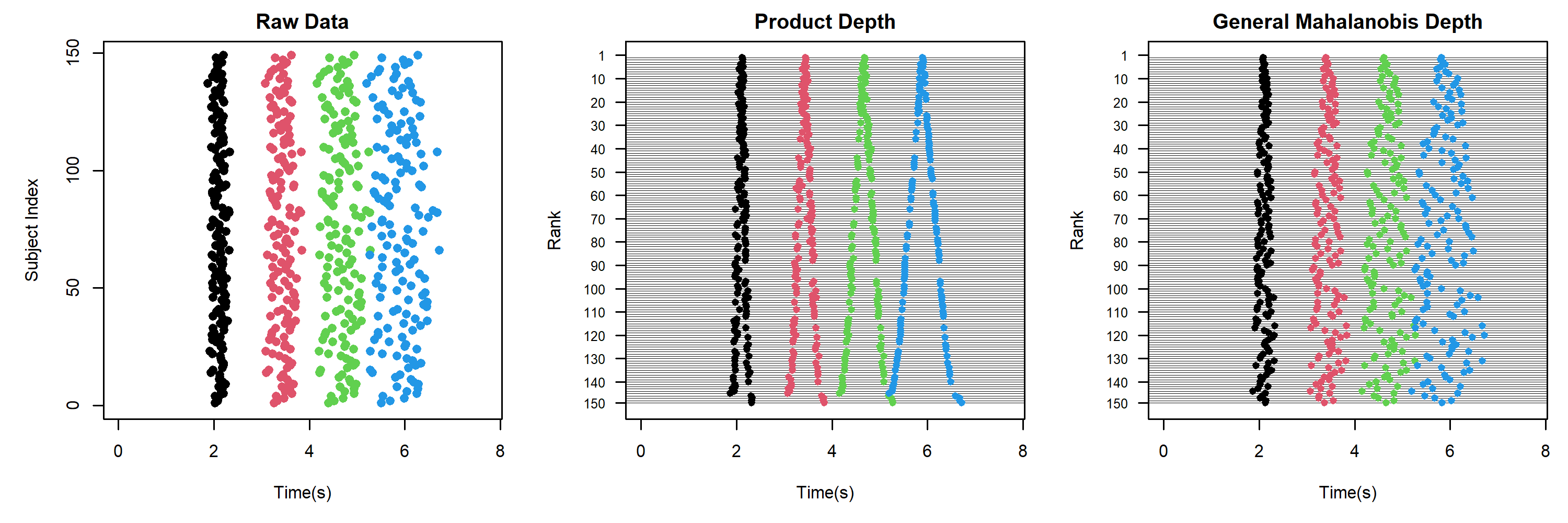}
      \caption{Left plot: Raw data; Middle plot: ranked data based on Product Depth; Right Plot: ranked data based on general Mahalanobis Depth. In all plots, black, red, green, and blue points correspond to the first, second, third and fourth event times ($S_1$, $S_2$, $S_3$ and $S_4$), respectively.}
      \label{fig:12}
\end{figure}

Figure \ref{fig:12} presents scatter plots of the raw data alongside the corresponding rankings derived from Product Depth and general Mahalanobis Depth. The left panels show the unranked raw data, while the middle and right panels display the data ranked according to Product Depth and Mahalanobis Depth, respectively. Compared with general Mahalanobis Depth, Product Depth produces a more structured and coherent ranking that aligns well with the temporal progression of sprint times. The center is more clearly identifiable, and the transitions across ranked subjects are smoother and more consistent.    

\section{Conclusion}
In this paper, we propose a novel depth function tailored for temporal point processes characterized by a fixed number of events over an unrestricted time domain. We investigate its mathematical properties, as well as ranking performance, through simulation studies. Compared to the general Mahalanobis Depth, the proposed Product Depth provides clearer notions of centrality and stronger within-group consistency. Applications on cell division data and 40m sprint test data also demonstrate its reasonable ranking to the TPPs observed only up to the first $k$ events. The proposed depth also shows potential for extension to tasks such as classification, clustering, and outlier detection.

\medskip
\noindent \textbf{Funding}: Dr. Shi's work was supported by the NSERC Discovery Grant RGPIN-2022-03264, the NSERC Alliance International Catalyst Grant ALLRP 590341-23, and the University of British Columbia Okanagan (UBC-O) Vice Principal Research in collaboration with UBC-O Irving K. Barber Faculty of Science. Dr. Fu’s research was supported by the NSERC Discovery Grant RGPIN 2018-05846. Dr. Chen’s research was supported by the NSERC Discovery Grant RGPIN 2022-04519.

\medskip
\noindent \textbf{Institutional Review Board Statement}: Not applicable.

\medskip
\noindent \textbf{Data Availability Statement:} Not applicable.

\medskip
\noindent \textbf{Conflicts of Interest:} The authors declare no conflict of interest.

\appendix
\section{Proof of Theorem \ref{thm1}}
\begin{itemize}
\item[$\mathbf{P1}$]$D(\mathbf{s}_k;P_{\mathbf{S}_k})$ is a continuous map from $\mathbb{S}_k$ to $\mathbb{R}^{+}$, and $D(\mathbf{s}_k;P_{\mathbf{S}_k})\to 0$ as $||\mathbf{s}_k|| \to \infty$, and  $D(\mathbf{s}_k;P_{\mathbf{S}_k})=0, \forall \ \mathbf{s}_k \in \mathbb{B}_k$. $\|\cdot\|$ denotes the $L_2$ (Euclidean) norm;
\item[$\mathbf{P2}$]  $D(\Theta_k;P_{\mathbf{S}_k})=\underset{\mathbf{s}_k \in \mathbb{S}_k}{\sup}D(\mathbf{s}_k;P_{\mathbf{S}_k})$ holds $\forall \ P_{\mathbf{S}_k} \in \mathbb{P}$ having central point $\Theta_k
$, where $\Theta_k= (\theta_1,\theta_2,\cdots,\theta_k)$, and $\theta_i=S_0+\mathbb{E}(S_k-S_0)\mathbb{E}\left(\sum^{i}_{j=1}\frac{S_j-S_{j-1}}{S_k-S_0}\right)$, $\forall i=1,2,\cdots,k$;
\item[$\mathbf{P3}$]
$\forall \ a \in \mathbb{R}^{+}$ and $\forall \  b \in \mathbb{R}$, $D(\mathbf{s}_k;P_{\mathbf{S}_k})=D(a\mathbf{s}_k+b;P_{a\mathbf{S}_k+b});$ and
\item[$\mathbf{P4}$]
$\forall \ P_{\mathbf{S}_k} \in \mathbb{P}$ having central point $\Theta_k,$ $D(\mathbf{s}_k;P_{\mathbf{S}_k}) \leq D\left(\Theta_k+\alpha\left(\mathbf{s}_k-\Theta_k\right);P_{\mathbf{S}_k}\right)$, $\forall \mathbf{s}_k \in \mathbb{S}_k, \alpha \in [0,1].$   
\end{itemize} 
We first prove $\mathbf{P2}$.
\begin{pf2}
   Since 
$$
\frac{\left[1+\frac{(s_k-s_0-\mu_{S_k-S_0})^2}{\sigma^2_{S_k-S_0}}\right]^{-1}}{\underset{t}{\max}\left\{\left[1+\frac{(t-s_0-\mu_{S_k-S_0})^2}{\sigma^2_{S_k-S_0}}\right]\right\}^{-1}} \leq 1,
$$
 and the equality holds when $s_k-s_0$ is the mean $\theta_k-s_0=\mu_{S_k-S_0}=\mathbb{E}(S_k-S_0)$ with respect to $P_{S_k}$. It is easy to see
$\omega(\boldsymbol{s}_k;P_{\boldsymbol{s}_k})^{\frac{|s_k-\eta|}{M-s_0}} \leq 1,$ and $\omega(\boldsymbol{s}_k;P_{\boldsymbol{s}_k})^{\frac{|s_k-\eta|}{M-s_0}}=1$ when $s_k-s_0=\theta_k-s_0.$

According to Young's inequality for products,
$$
\prod a_i^{p_i} \leq \sum_{i} p_ia_i
$$
if $0\leq p_i \leq 1$ with $\sum_ip_i=1, \forall a_i \geq 0, i=1,2,\cdots.$
\begin{eqnarray*}
    \prod_{i=1}^{k}\left[\frac{s_i-s_{i-1}}{(s_k-s_0)\bar{\mu}_i}\right]^{\bar{\mu}_i}
    &\leq& \sum_{i=1}^k \bar{\mu}_i\left[\frac{s_i-s_{i-1}}{(s_k-s_0)\bar{\mu}_i}\right]\\
    &=& \sum_{i=1}^{k} \left[\frac{s_i-s_{i-1}}{(s_k-s_0)}\right]\\
    &=& 1,
\end{eqnarray*}
The equality holds when $\frac{s_i-s_{i-1}}{s_k-s_0}=\bar{\mu}_i=\mathbb{E}(\frac{S_i-S_{i-1}}{S_k-S_0}).$ Thus, the depth value of the central point, $\Theta_k$ is maximum. 
\end{pf2}

\begin{pf1}
It is easy to see the continuity and vanishing at the boundary for the defined depth function. 
Since $||\mathbf{s}_k|| \to \infty$ implies $s_k \to \infty,$ $\omega(s_k,P_{S_k}) \to 0$ as $||\mathbf{s}_k|| \to \infty.$ This implies  $\omega(s_k; P_{S_k})^{\frac{|s_k-\eta|}{M-s_0}} \to 0$ as $||\mathbf{s}_k|| \to \infty.$ Since the conditional depth is bounded, the depth function vanishes at infinity. 
\end{pf1}

\begin{pf3}
Let $a \in \mathbb{R}^{+}$ and $b \in \mathbb{R}.$ 
Since $\left[1+\frac{\left(s_k-s_0-\mu_{S_k-S_0}\right)^2}{\sigma^2_{S_k-S_0}}\right]^{-1}=\left[1+\frac{\left(a(s_k-s_0)-a\mu_{S_k-S_0}\right)^2}{a^2\sigma^2_{S_k-S_0}}\right]^{-1}$,  
$$
\omega(s_k; P_{S_k}) = \omega(as_k+b; P_{aS_k+b}).
$$
Since $M$ is the mean $\mathbb{E}(S_k)$ with respect to $P_{S_k}$ in this paper,
$$
M(P_{aS_k+b}) = aM(P_{S_k})+b.
$$
Similar, 
$$
\eta(P_{aS_k+b}) = a\eta(P_{S_k})+b.
$$
Since $\forall i=1,2,\cdots k,$
$$
\frac{|s_k-\eta(P_{S_k})|}{M(P_{S_k})-s_0}=\frac{|as_k+b-a\eta(P_{S_k})-b|}{aM(P_{S_k})+b-as_0-b}=\frac{|(as_k+b)-\eta(P_{aS_k+b})|}{M(P_{aS_k+b})-(as_0+b)},
$$
Thus,
$
\omega(s_k; P_{S_k})^{\frac{|s_k-\eta|}{M-s_0}} \text{ holds the property.}
$\\
It is easy to see that the conditional depth holds the property.\\
Therefore,
$$
D(\boldsymbol{s}_k;P_k)=D(a\boldsymbol{s}_k+b;P_{a\boldsymbol{S}_k+b}).
$$
\end{pf3}

To prove $\mathbf{P4}$, we need to introduce Lemma \ref{lem1} and Lemma \ref{lem2} in advance.

\begin{lem}\label{lem1}
Let $f_{\Theta_k,\mathbf{s}_k}(\alpha)=D\left(\Theta_k+\alpha(\mathbf{s}_k-\Theta_k);P_{\mathbf{S}_k}\right),$ where $\alpha \in [0,1].$ Then, $\forall \  \mathbf{s}_k \in \mathbb{S}_k,\ 
\forall \  \alpha \in [0,1],$
\begin{equation*}
f_{\Theta_k,\mathbf{s}_k}(1)\leq f_{\Theta_k,\mathbf{s}_k}(\alpha) \iff f_{\Theta_k,\mathbf{s}_k}(\alpha) \textit{is a decreasing function in}\  \alpha.    
\end{equation*}
\end{lem}

\begin{lem}\label{lem2}
For $(a_1,a_2,\cdots,a_k)$, and $(b_1,b_2,\cdots,b_k),$ where $k\in \mathbb{N}^+$, $\alpha \in [0,1]$, and $a_i \geq 0$, $b_i \geq 0$, for $i=1,2,\cdots, k$,
$$
\sum_{i=1}^{k}\frac{a_i-b_i}{(1-\alpha)+\alpha\frac{a_i}{b_i}} \leq \frac{\sum_{i=1}^{k}a_i-\sum_{i=1}^{k}b_i}{(1-\alpha)+\alpha\frac{\sum_{i=1}^{k}a_i}{\sum_{i=1}^{k}b_i}}.
$$
\end{lem}
\begin{pol1}
We aim to show that the depth value of $\mathbf{s}_{k}$ decreases monotonically as it moves away from the center, $\Theta_{k}$, along any fixed ray through the center. This is equivalent to showing that 
$f_{\Theta_k,\mathbf{s}_k}(\alpha)$ is decreasing in $\alpha$ for all 
$\mathbf{s}_k \in \mathbb{S}_k$ and all $\alpha \in [0,1]$.
\begin{equation}
\forall \  \mathbf{s}_k \in \mathbb{S}_k,\ 
\forall \  \alpha \in [0,1], \ 
f_{\Theta_k,\mathbf{s}_k}(1)\leq f_{\Theta_k,\mathbf{s}_k}(\alpha).    
\end{equation}
$\iff$
\begin{equation}
f_{\Theta_k,\mathbf{s}_k}(\alpha) \textit{ is a decreasing function in}\  \alpha, \  \forall \mathbf{s}_k \in \mathbb{S}_k,\ 
\alpha \in [0,1].
\end{equation}
(A.2) $\implies$ (A.1)
$$
\forall \ 0\leq \alpha_1 \leq \alpha_2 \leq 1, \ f_{\Theta_k,\mathbf{s}_k}(\alpha_2) \leq f_{\Theta_k,\mathbf{s}_k}(\alpha_1),
$$
$\textit{let} \  \alpha_1=\alpha, \ \alpha_2=1,$
$$
f_{\Theta_k,\mathbf{s}_k}(1)\leq f_{\Theta_k,\mathbf{s}_k}(\alpha).
$$
(A.1)$\implies$ (A.2)

$$\forall \  \mathbf{y}_k\in \mathbb{S}_k, \alpha_1\in [0,1], \alpha\in [0,1], \  \ \text{let} \ \mathbf{s}_k =\Theta_k+\alpha_1(\mathbf{y}_k-\Theta_k).
$$
Since $\mathbf{s}_k \in \mathbb{S}_k,$
\begin{eqnarray*}
    f_{\Theta_k,\mathbf{s}_k}(1)
    &=&D\left(\mathbf{s}_k;P_{\mathbf{S}_k}\right)\\
    &=& D\left(\Theta_k+\alpha_1(\mathbf{y}_k-\Theta_k);P_{\mathbf{S}_k}\right)\\
    &\leq& D\left(\Theta_k+\alpha\left(\Theta_k+\alpha_1(\mathbf{y}_k-\Theta_k)-\Theta_k\right)\right)\\
    &=& D\left(\Theta_k+\alpha_1\alpha(\mathbf{y}_k-\Theta_k);P_{\mathbf{S}_k}\right)\\
    &=&
    f_{\Theta_k,\mathbf{s}_k}(\alpha_1\alpha).\\
\end{eqnarray*}
$\implies$ 
$$
f_{\Theta_k\mathbf{y}_k}(\alpha_1)\leq f_{\Theta_k\mathbf{y}_k}(\alpha_1\alpha).
$$
Since $\alpha_1 \geq \alpha_1\alpha,$ $f_{\Theta_k\mathbf{y}_k}(\alpha)$ is a decreasing function in $\alpha$, $\forall \mathbf{y}_k \in \mathbb{S}_k,\ 
\alpha \in [0,1].$
\end{pol1}

\begin{pol2}
To show that $f_{\Theta_k,\mathbf{s}_k}(\alpha)$ is decreasing in $\alpha$ for all 
$\mathbf{s}_k \in \mathbb{S}_k$ and all $\alpha \in [0,1]$, we need to prove Lemma \ref{lem2}.\\
When $k= 1$, it is trivial.\\
When $k= 2$,
$\forall a_1 > 0$, $a_2> 0$, $b_1>0$, $b_2> 0$, $\alpha \in [0,1],$ let
$$
W=\frac{a_1-b_1}{(1-\alpha)+\alpha\frac{a_1}{b_1}}
+\frac{a_2-b_2}{(1-\alpha)+\alpha\frac{a_2}{b_2}}
- \frac{(a_1+a_2)-(b_1+b_2)}{(1-\alpha)+\alpha\frac{a_1+a_2}{b_1+b_2}}\hspace{2cm}
$$
Let $A_1=(1-\alpha)b_1+\alpha a_1$,$A_2=(1-\alpha)b_2+\alpha a_2,$ $C =(1-\alpha)(b_1+b_2)+\alpha (a_1+a_2),$ 
\begin{align*}
\begin{split}
W= \frac{(a_1-b_1)b_1A_2C+(a_2-b_2)b_2A_1C-(a_1+a_2-b_1-b_2)(b_1+b_2)A_1A_2}{A_1A_2C}\\
=\frac{(a_1-b_1)A_2\left[b_1C-A_1(b_1+b_2)\right]+(a_2-b_2)A_1\left[b_2C-A_2(b_1+b_2)\right]}{A_1A_2C}.\hspace{0.4cm}
\end{split} 
\end{align*}
Since 
\begin{eqnarray*}
    b_1C-A_1(b_1+b_2)
    &=& b_1(C-A_1)-b_2A_1\\
    &=& \alpha(a_2b_1-a_1b_2),
\end{eqnarray*}
and similarly, 
\begin{eqnarray*}
    b_2C-A_2(b_1+b_2)
    &=& \alpha(a_1b_2-a_2b_1),
\end{eqnarray*}

\begin{align*}
\begin{split}
W=\frac{(a_1-b_1)A_2\alpha(a_2b_1-a_1b_2)+(a_2-b_2)A_1\alpha(a_1b_2-a_2b_1)}{A_1A_2C}\\
=\frac{\left[(a_1-b_1)A_2-(a_2-b_2)A_1\right]\alpha(a_2b_1-a_1b_2)}{A_1A_2C}.\hspace{1.8cm}
\end{split}
\end{align*}
Since
\begin{align*}
\begin{split}
    (a_1-b_1)A_2-(a_2-b_2)A_1=\left[(a_1-b_1)\left[(1-\alpha)b_2+\alpha a_2\right]-(a_2-b_2)\left[(1-\alpha)b_1+\alpha a_1\right]\right]\\
    =a_1b_2-a_2b_1, \ \ \ \ \ \ \ \ \ \ \ \ \ \ \ \ \ \ \ \ \ \ \ \ \ \ \ \ \ \ \ \ \ \ \ \ \ \ \ \ \ \ \ \ \ \ \ \ \ \ \ \ \ \ \ \ \ \ \ \ 
\end{split}
\end{align*}
$$
W=\frac{-\alpha(a_1b_2-a_2b_1)^2}{A_1A_2C}.
$$
Since $A_1A_2C >0$, $W \leq 0.$\\\
Assume the inequality holds when $k \leq n, \ \forall n \geq 2$, then for $k=n+1$,
\begin{eqnarray*}
    \sum_{i=1}^{n+1}\frac{a_i-b_i}{(1-\alpha)+\alpha\frac{a_i}{b_i}}
    &=& \sum_{i=1}^{n}\frac{a_i-b_i}{(1-\alpha)+\alpha\frac{a_i}{b_i}}
    +\frac{a_{n+1}-b_{n+1}}{(1-\alpha)+\alpha\frac{a_{n+1}}{b_{n+1}}}\\
    &\leq& \frac{\sum_{i=1}^{n}a_i-\sum_{i=1}^{n}b_i}{(1-\alpha)+\alpha\frac{\sum_{i=1}^{n}a_i}{\sum_{i=1}^{n}b_i}}
    +\frac{a_{n+1}-b_{n+1}}{(1-\alpha)+\alpha\frac{a_{n+1}}{b_{n+1}}}\\
    &\leq& \frac{\sum_{i=1}^{n+1}a_i-\sum_{i=1}^{n+1}b_i}{(1-\alpha)+\alpha\frac{\sum_{i=1}^{n+1}a_i}{\sum_{i=1}^{n+1}b_i}}.
\end{eqnarray*}
Thus, the inequality holds.
\end{pol2}

With Lemma 1 and Lemma 2, the property P-4 can be proved as follows.

\begin{pf4}
The depth function is defined as 
$$
D(\mathbf{s}_k;P_{\mathbf{S}_k})=\left[\omega(s_k;P_{S_k})\right]^{\frac{|s_k-\eta|}{M-s_0}}\prod_{i=1}^{k}\left[\frac{s_i-s_{i-1}}{(s_k-s_0)\bar{\mu}_i}\right]^{\bar{\mu}_i},
$$
where $\eta=\underset{t}{\arg\max}\ \left(\omega(t;P_{S_k})\right)$, $M=\mathbb{E}(S_k)$ with respect to $P_{S_k}$, $\bar{\mu}_i=\frac{\theta_i-\theta_{i-1}}{\theta_k-\theta_0},$ for $i=1,2,\cdots, k$, and $\theta_0=s_0=T_0$.\\
Let $\mathbb{B}_k=\{(s_1,s_2,\cdots,s_k)) \in \mathbb{S}_k|\textit{at least one equality holds}\}$ be the boundary, the depth equals $0$ at the boundary. Thus, we only consider $\mathbf{s}_k \in \mathbb{S}_k\setminus \mathbb{B}_k$ when we prove the property of monotonically decreasing from the center.\\
Let $\mathbf{s}_k \in \mathbb{S}_k\setminus \mathbb{B}_k,$ and $\alpha \in [0,1]$ be arbitrary,
denote 
$$
H_{\Theta_k,\mathbf{s}_k}(\alpha) = D\left(\Theta_k+\alpha(\mathbf{s}_k-\Theta_k);P_{\mathbf{S}_k}\right),
$$ 
$$
f_{\Theta_k,\mathbf{s}_k}(\alpha)
= D_c\left(\Theta_k+\alpha(\mathbf{s}_k-\Theta_k);P_{\mathbf{S}_k|S_k}\right), 
$$
$$
h_{\theta_k,s_k}(\alpha)=\left[\omega\left(\theta_k+\alpha(s_k-\theta_k);P_{S_k}\right)\right]^{\frac{|\theta_k+\alpha\left(s_k-\theta_k\right)-\eta|}{M-s_0}}.
$$
With Lemma1, it is equivalent to proving $H_{\Theta_k,\mathbf{s}_k}(\alpha) =h_{\theta_k,x_k}(\alpha)f_{\Theta_k,\mathbf{s}_k}(\alpha)$ is a decreasing function of $\alpha.$ Since $h_{\theta_k,x_k}(\alpha) \geq 0$ and $f_{\Theta_k,\mathbf{s}_k}(\alpha) \geq 0$, if we can prove both of them are decreasing functions in $\alpha$, then we can see that the depth function satisfies the property.\\
We first explore the conditional depth,
$$
D_c(\mathbf{s}_k;P_{\mathbf{S}_k|S_k})=\prod_{i=1}^{k}\left(\frac{s_i-s_{i-1}}{(s_k-s_0)\bar{\mu}_i}\right)^{\bar{\mu}_i}.
$$
\begin{eqnarray*}
f_{\Theta_k,\mathbf{s}_k}(\alpha)
&=& \prod_{i=1}^{k}\left[\frac{(\theta_i-\theta_{i-1})+\alpha\left[(s_i-s_{i-1})-(\theta_i-\theta_{i-1})\right]}{\left[\theta_k-s_0+\alpha(s_k-\theta_k)\right]\bar{\mu}_i}\right]^{\bar{\mu}_i}.\\
\end{eqnarray*}
Denote
\begin{eqnarray*}
g_{\Theta_k,\mathbf{s}_k}(\alpha)
&=& \log f_{\Theta_k,\mathbf{s}_k}(\alpha)\\
&=& \sum_{i=1}^{k}\left[\bar{\mu}_i\log \left[(1-\alpha)(\theta_i-\theta_{i-1})+\alpha (s_i-s_{i-1})\right]\right]\\
&-& \sum_{i=1}^{k}\left[\bar{\mu}_i\log \left[(\theta_k-s_0)+\alpha (s_k-\theta_k)\right]\right]-
\sum_{i=1}^{k}\bar{\mu}_i\log\bar{\mu}_i \\
&=& \sum_{i=1}^{k}\left[\bar{\mu}_i\log \left[(1-\alpha)(\theta_i-\theta_{i-1})+\alpha (s_i-s_{i-1})\right]\right]\\
&-& \log \left[(\theta_k-s_0)+\alpha (s_k-\theta_k)\right]-\sum_{i=1}^{k}\bar{\mu}_i\log\bar{\mu}_i. \\
\end{eqnarray*}
The first derivative of $g_{\Theta_k,\mathbf{s}_k}(\alpha)$ with respect to $\alpha$ is as follows
\begin{eqnarray*}
g'_{\Theta_k,\mathbf{s}_k}(\alpha)
&=& \sum_{i=1}^{k}\bar{\mu}_i
\frac{(s_i-s_{i-1})-(\theta_i-\theta_{i-1})}
{(1-\alpha)(\theta_i-\theta_{i-1})+\alpha(s_i-s_{i-1})}
-\frac{s_k-\theta_k}{(\theta_k-s_0)+\alpha(s_k-\theta_k)}\\
&=& \sum_{i=1}^{k}
\frac{\frac{s_i-s_{i-1}}{\theta_k-\theta_0}-\frac{\theta_i-\theta_{i-1}}{\theta_k-\theta_0}}
{(1-\alpha)+\alpha\frac{s_i-s_{i-1}}{\theta_i-\theta_{i-1}}}
-\frac{\frac{s_k-\theta_k}{\theta_k-\theta_0}}
{(1-\alpha)+\alpha\frac{s_k-s_0}{\theta_k-\theta_0}}
\qquad (\text{since } s_0=\theta_0=T_0)\\
&=& \frac{1}{\theta_k-\theta_0}\left[
\sum_{i=1}^{k}
\frac{(s_i-s_{i-1})-(\theta_i-\theta_{i-1})}
{(1-\alpha)+\alpha\frac{s_i-s_{i-1}}{\theta_i-\theta_{i-1}}}
-\frac{s_k-\theta_k}
{(1-\alpha)+\alpha\frac{s_k-s_0}{\theta_k-\theta_0}}
\right].
\end{eqnarray*}
According to the Lemma 2, $\sum_{i=1}^{k}\frac{(s_i-s_{i-1})-(\theta_i-\theta_{i-1})}{(1-\alpha)+\alpha\frac{s_i-s_{i-1}}{\theta_i-\theta_{i-1}}} \leq \frac{s_k-\theta_k}{(1-\alpha)+\alpha\frac{s_k-s_0}{\theta_k-\theta_0}}.$
Thus, $g_{\Theta_k,\mathbf{s}_k}^{'}(\alpha) \leq 0$, the function $g_{\Theta_k,\mathbf{s}_k} (\alpha)$ is a decreasing function in $\alpha$. This implies $f_{\Theta_k,\mathbf{s}_k}(\alpha)$ is a decreasing function in $\alpha$.\\
Secondly, we explore the $\omega$ part
$$
\omega(s_k; P_{S_k})^{\frac{|s_k-\eta|}{M-s_0}},
$$
where $\eta=\underset{t}{\arg\max}\ \left(\omega(t;P_{S_k})\right).$\\
Since $\omega(s_k; P_{S_k}) \leq 1$,\ $\omega(s_k; P_{S_k})^{\beta} \leq 1, \  \forall \ \beta \geq 0.$ \\
Let $\theta_k = \underset{t}{\arg\max}\ \left(\omega(t; P_{S_k})\right)=\eta$,
$$
\omega(\theta_k; P_{S_k})^{\frac{|s_k-\theta_k|}{M-s_0}}=1=\underset{t}{\max}\ \left(\omega(t; P_{S_k})\right)^{\frac{|t-\eta|}{M-s_0}},
$$ 
where $\eta=\underset{t}{\arg\max}\ \left(\omega(t; P_{S_k})\right).$ Thus, $\theta_k$ is the center of $S_k.$\\
Denote 
$$
h_{\theta_k,s_k}^{(1)}(\alpha) =\omega(\theta_k+\alpha(s_k-\theta_k); P_{S_k}),
$$
\begin{eqnarray*}
h_{\theta_k, s_k}^{(2)}(\alpha)
&=&\frac{|\theta_k+\alpha(s_k-\theta_k)-\eta|}{M}\\
&=& \frac{\alpha|(s_k-\theta_k)|}{M-s_0}.
\end{eqnarray*}
Then
$$
h_{\theta_k,s_k}(\alpha) = h_{\theta_k,s_k}^{(1)}(\alpha)^{h_{\theta_k,s_k}^{(2)}(\alpha)}.
$$
The $\omega\big(\theta_k+\alpha(s_k-\theta_k)\big)$ we choose is monotonically decreasing from the center, $\theta_k.$ Due to Lemma 1, $h_{\theta_k,s_k}^{(1)}(\alpha)$ is a decreasing function of $\alpha, $ and $\underset{\alpha}{\max }\ h_{\theta_k,s_k}^{(1)}(\alpha) = 1.$ Obviously, $h_{\theta_k,s_k}^{(2)}(\alpha)$ is a increasing function in $\alpha$ with $\underset{\alpha}{\min}\ h_{\theta_k,s_k}^{(2)}(\alpha)=0.$ 
Thus, $\log h_{\theta_k,s_k}(\alpha) =h_{\theta_k,s_k}^{(2)}(\alpha)\log h_{\theta_k,s_k}^{(1)}(\alpha)$ is a decreasing function in $\alpha$, which implies that $h_{\theta_k,s_k}(\alpha)$ is a decreasing function in $\alpha$.  

Since both $f_{\Theta_k,\mathbf{s}_k}(\alpha)$ and $h_{\theta_k,s_k}(\alpha)$ are decreasing functions of $\alpha,$ $H_{\Theta_k,\mathbf{s}_k}(\alpha)$ is a decreasing function in $\alpha$. Therefore, the depth function satisfies the property of monotonically decreasing from the center.
\end{pf4}

\end{document}